\documentclass[%
 reprint,
superscriptaddress,
 amsmath,amssymb,
 aps,
 pra,
floatfix
]{revtex4-2}

\usepackage{graphicx}
\usepackage{dcolumn}
\usepackage{bm}
\usepackage{hyperref}

\usepackage{braket}
\usepackage{xfrac}
\usepackage{siunitx}
\DeclareSIUnit\gauss{G}
\usepackage[english]{babel}

\begin{document}

\title{Realization of topological Thouless pumping in a synthetic Rydberg dimension}

\author{Martin Trautmann}
 \thanks{Present address: Max Planck Institute of Microstructure Physics, D-06120 Halle}
 \affiliation{%
 Felix-Bloch Institute, Leipzig University, Linnéstraße 5, 04103 Leipzig, Germany
}%
\author{Inti Sodemann Villadiego}%
 \email{sodemann@uni-leipzig.de}
\affiliation{%
 Institute for theoretical physics, Leipzig University, Brüderstraße 16, 04103, Leipzig, Germany
}%
\author{Johannes Deiglmayr}%
 \email{johannes.deiglmayr@uni-leipzig.de}
\affiliation{%
  Felix-Bloch Institute, Leipzig University, Linnéstraße 5, 04103 Leipzig, Germany
}%

\date{\today}

\begin{abstract}
    The simulation of synthetic dimensions by manipulating internal states of atoms and molecules has opened the door to investigate regimes outside those of more traditional quantum many-body platforms. Highly excited \textit{Rydberg states} of atoms are a particularly promising platform to engineer Hamiltonians in such synthetic dimensions due to their large number of addressable states and the readily available technologies for manipulating their couplings and for detecting them. In this letter, we demonstrate the realization of topological quantum pumping in synthetic dimensions by engineering a one-dimensional Rice-Mele chain from the Rydberg states of cesium atoms, and manipulating their couplings in a time-dependent fashion through radio-frequency fields. We implement Thouless protocols for topological pumping and investigate the efficiency for pumping an effective quantum particle as a function of the period of pumping and other parameters while allowing for rates of change that are not necessarily adiabatic. We demonstrate that optimal pumping efficiencies of up to \SI{90}{\percent} can be achieved when the pump is operated in the topological Thouless regime, even when the pumping is accompanied by the wave-packet spread that arises from the energy dispersion of the particle along the synthetic dimension.
\end{abstract}

\maketitle


 The concept of synthetic dimensions has been introduced in quantum simulations to encode the translational motion along a discretized spatial dimension into the internal states of atoms or molecules\cite{boadaQuantumSimulationExtra2012,celiSyntheticGaugeFields2014a}. This concept has been applied, \textit{e.g.}, to the study of chiral edge states in the quantum Hall regime, the prototypical realization of topological effects in solid-state systems~\cite{vonklitzingQuantizedHallEffect1986}, by coupling the spin states of ultracold ground-state atoms through Raman transitions~\cite{manciniObservationChiralEdge2015,stuhlVisualizingEdgeStates2015} or by coupling arrays of photonic waveguides ~\cite{zilberbergPhotonicTopologicalBoundary2018}. A great asset of this approach is the possibility to engineer synthetic gauge fields~\cite{dalibardColloquiumArtificialGauge2011,goldmanLightinducedGaugeFields2014a}. Recently it has been realized that coupling the rotational states of molecules~\cite{sundarSyntheticDimensionsUltracold2018} or highly-excited Rydberg states of atoms~\cite{ozawaTopologicalQuantumMatter2019a} via radio-frequency (rf) fields offers the possibility to significantly increase the size of the synthetic dimensions and the controllability of interactions. Rf-coupled Rydberg states have been used to study the appearance of topologically protected edge states in the Su-Schrieffer-Heeger model~\cite{suSolitonsPolyacetylene1979,kanungoRealizingTopologicalEdge2022,luWavepacketDynamicsLongrange2024} or to realize artificial gauge fields~\cite{chenStronglyInteractingRydberg2024}. However, in these seminal studies, only time-independent Hamiltonians were implemented. The quantum simulation of explicitly time-dependent Hamiltonians with dynamically varying parameters is of great interest because they can allow to engineer a variety of many-body dynamical regimes, such as the Floquet engineering of topological many-body states \cite{okaFloquetEngineeringQuantum2019,cayssolFloquetTopologicalInsulators2013,rodriguez-vegaLowfrequencyMoireFloquet2021}.

\begin{figure}[tb]
	\centering
    \includegraphics[width=\linewidth]{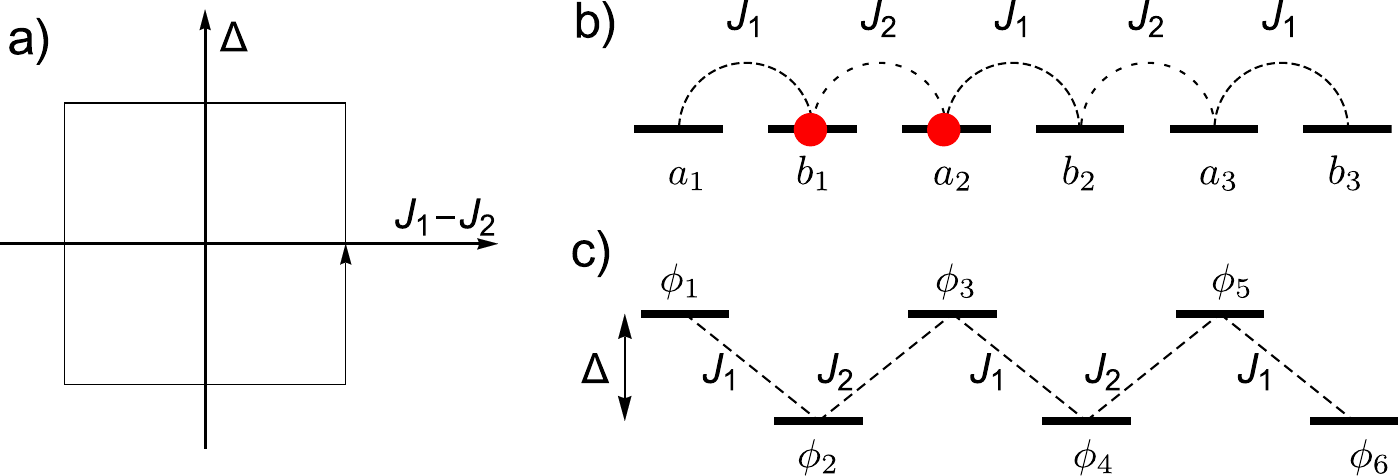}
	\caption{\label{fig:SOAT_schematic_thouless} a) Depiction of a closed loop in the parameter space of the Rice-Mele (RM) Hamiltonian (Eq.~\eqref{eq:SOAT_rice-mele}) which encloses the origin where the band gap closes. b) Illustration of a six-sites RM system with vanishing on-site energies. Red circles mark the initial superposition of sites that was prepared to demonstrate quantum transport (see text for details). c) Illustration of a six-sites RM system with an imbalance $\Delta$ of on-site energies and the mapping to our physical Rydberg states $\phi_i$.}
\end{figure}

  In this study we will investigate a paradigmatic example of the interplay of topology and Hamiltonian time dependence, first pointed out by Thouless~\cite{thoulessQuantizationParticleTransport1983}, namely that the charge pumped over one period in a one-dimensional gapped system is quantized. Thouless pumping has been observed in numerous systems~\cite{citroThoulessPumpingTopology2023}, such as artifical spin systems~\cite{schroerMeasuringTopologicalTransition2014} or neutral atoms in optical lattices~\cite{nakajimaTopologicalThoulessPumping2016,lohseThoulessQuantumPump2016}. Here we demonstrate Thouless pumping in a Rice-Mele (RM) model, realized in a Rydberg synthetic dimension in cesium atoms. The RM model describes a chain of dimers consisting of atoms $a$ and $b$~\cite{riceElementaryExcitationsLinearly1982}
\begin{eqnarray}\label{eq:SOAT_rice-mele}
  \hat{H}_\mathrm{RM}(t)=& -\sum_{j=1}^N \left( J_1(t)\hat{b}_j^\dagger \hat{a}_j^{\,} + J_2(t)\hat{a}_{j+1}^\dagger \hat{b}_{j}^{\,} + h.c. \right) + \nonumber \\
  & \Delta(t) \sum_{j=1}^N \left( \hat{a}_j^\dagger \hat{a}_j^{\,}-\hat{b}_j^\dagger \hat{b}_j^{\,}\right),
\end{eqnarray}
where $J_1\left(t\right)$ and $J_2\left(t\right)$ are the time-dependent couplings within a dimer, and between neighboring dimers, respectively, and $\Delta\left(t\right)$ is the time-dependent imbalance of on-site energies within a dimer. The operators $\hat{a}_j^\dagger,\hat{a}_j$ and $\hat{b}_j^\dagger,\hat{b}_j$ are creation and annihilation operators at site $a$ and $b$ of the dimer $j$. We consider the case where the parameters $J_1\left(t\right)$, $J_2\left(t\right)$ and $\Delta\left(t\right)$ are modulated periodically in time such that $J_1 - J_2$ and $\Delta$ follow the trajectory shown in Fig.~\ref{fig:SOAT_schematic_thouless}\,a), which is a realization of the Thouless pump (see appendix~\ref{sec:optimalperiod}).

 In order to implement the Rice-Mele model in a synthetic dimension, up to six Rydberg states of cesium are coupled by rf fields, as depicted in Figure~\ref{fig:SOAT_schematic_thouless}\,b) and c). The coupled system is described in the rotating-wave approximation by the Hamiltonian~\eqref{eq:SOAT_rice-mele} where the couplings $J_{1/2}$ map onto Rabi frequencies $\Omega_{1/2}$, and the imbalance of on-site energies $\Delta$ is created by detuning the rf fields~\cite{kanungoRealizingTopologicalEdge2022}. Specifically, we couple the Rydberg states $\phi_1=\ket{n=55,l=0,j=1/2,m_j=1/2}$, $\phi_2=\ket{55,1,3/2,3/2}$, $\phi_3=\ket{56,0,1/2,1/2}$, $\phi_4=\ket{56,1,3/2,3/2}$, $\phi_5=\ket{57,0,1/2,1/2}$, and $\phi_6=\ket{57,1,3/2,3/2}$ sequentially by near-resonant rf fields. The basis states have been chosen on the basis of two principles: \textit{i)} the selected transitions are among the strongest electric dipole transitions between Rydberg states and have comparable transition dipole moments. This minimizes the effect of off-resonant couplings to other Rydberg states, which cause rf-power dependent ac Stark shifts; \textit{ii)} the transition frequencies are similar enough so that they can be created and applied in a single rf setup, while at the same time the differences in the transition frequencies are large enough so that each coupling can be controlled separately. The transition frequencies range from \SI{21.5}{\giga\hertz} to \SI{24.3}{\giga\hertz} (with the largest deviation from the mean being around \SI{7}{\percent}) and all differ by more than \SI{250}{\mega\hertz} from each other which is significantly larger than the Rabi frequencies and detunings explored in this work.

 \begin{figure}[tb]
	\centering
        \includegraphics[width=\linewidth]{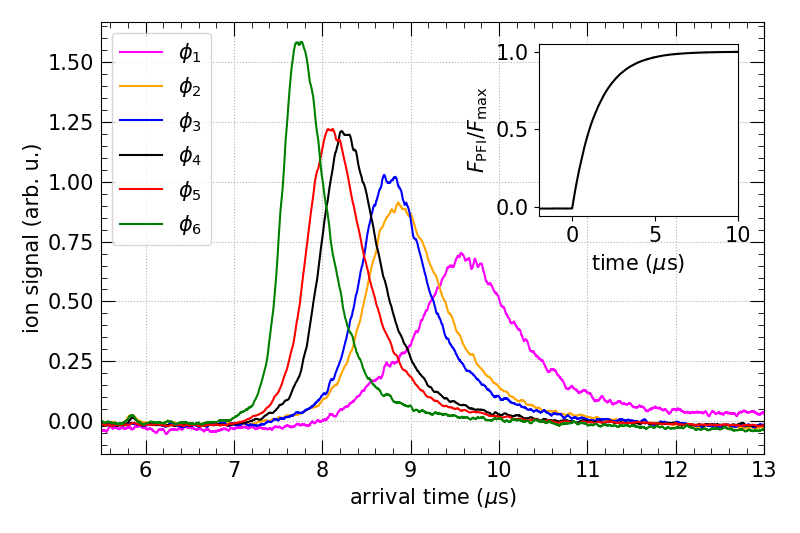}
    	\caption{\label{fig:SOAT_exp_tof_traces} Currents measured at the micro-channel-plate detector (area-normalized) after ionization of atoms in different Rydberg states by applying a ramped electric field $F_\mathrm{PFI}$ (see inset, $F_\mathrm{max}\simeq\SI{70}{\volt\per\centi\meter}$). States $\{\phi_2,\phi_4,\phi_6\}$ ($\{\phi_1,\phi_3,\phi_5\}$) were excited by direct one-photon (off-resonant two-photon UV-rf) excitation. The inset shows the measured electric potential applied to the electrodes ($U_\mathrm{max}=\SI{2.2}{\kilo\volt}$). The shape results from the insertion of a low-pass filter after a high-voltage pulser.}
\end{figure}

 The five fields are created by frequency-doubling the output of a single arbitrary waveform generator (Tektronix AWG70001A-150-AC) using a passive doubler (Mini-Circuits ZXF90-2-44-K+) and are coupled to free space by a pyramidal horn antenna (RF spin H-A40). The horn is placed directly in front of a fused-silica viewport of the UHV chamber with an open diameter of \SI{38}{\mm}, which reduces the effects of diffraction from this circular aperture. The resonant transition frequencies are calculated from accurate quantum defects~\cite{deiglmayrPrecisionMeasurementIonization2016} and are verified by performing rf spectroscopy of the transitions with single, weak rf tones. The details of the setup for high-resolution spectroscopy of Rydberg states has been described previously~\cite{peperPrecisionMeasurementIonization2019}.

 \textit{Site-selective detection.} Populations in the different Rydberg states are measured by state-selective pulse-field ionization using ramped electric fields~\cite{fabreMillimetreResonancesNa1977}. This allows one to measure directly the probability amplitude of the system's wave function at each site in the eigenstate basis. The technique of state-sensitive detection is based on the dependence of the ionization threshold on the (effective) quantum number $n$, \textit{e.g.}, for atoms in states with low angular momentum the classical threshold field $F_\mathrm{ion}$ scales as $F_\mathrm{ion} = \frac{E_\mathrm{h}}{e a_0}\frac{1}{16 n^4}$. In a rising electric field pulse, energetically higher-lying Rydberg states ionize earlier and the resulting ions arrive earlier at the detector. This trend is clearly visible in Figure~\ref{fig:SOAT_exp_tof_traces}, which depicts experimental time-of-flight traces recorded for the field ionization of the chosen Rydberg states in identically ramped electric fields. The quantum-state-dependent ionization thresholds thus yield distinct distributions of the arrival times of ions at the detector.

\begin{figure}[tb]
        \includegraphics[width=0.85\linewidth]{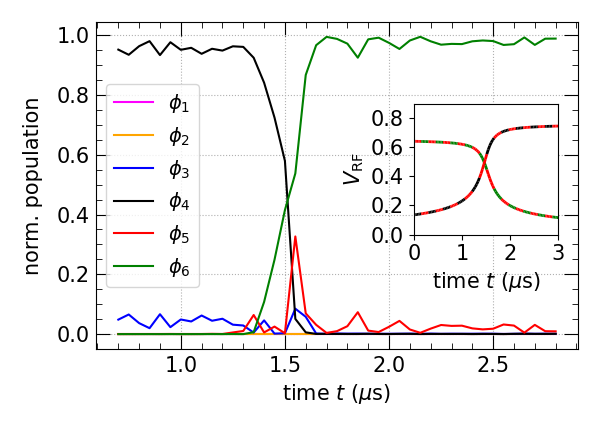}
	\caption{\label{fig:SOAT_TIRAP} Populations during a STIRAP sequence for transfer from $\phi_4$ to $\phi_6$. The populations are determined by fitting the measured current traces to a weighted sum of the basis traces shown in Fig.~\ref{fig:SOAT_exp_tof_traces} (calibrated peak Rabi frequency for both transitions $\Omega= 2 \pi \times \SI{8.5}{\mega\hertz}$). Inset: amplitudes of the programmed waveforms coupling $\phi_4\rightarrow \phi_5$ (black-red--dashed) and $\phi_5\rightarrow \phi_6$ (red-green--dashed). Note that the amplitude of the rf field applied to the atoms is proportional to the square of the programmed amplitudes because of the use of a frequency doubler.}
\end{figure}

 Because the ion-arrival-time distributions of the different field-ionized basis states overlap, see Fig.~\ref{fig:SOAT_exp_tof_traces}, additional analysis is required to extract the population of basis states from the experimental data. The time-of-flight trace resulting from ionization of a Rydberg atom in a superposition $\psi=\sum_i c_i \phi_i$ of the basis states is a weighted sum of the state-specific traces where the weights are proportional to the populations $\left|c_i\right|^2$. We extract the relative populations using a non-linear optimization routine with the traces depicted in Fig.~\ref{fig:SOAT_exp_tof_traces} as basis functions. To test the reliability of the method, we performed measurements such as STIRAP (stimulated Raman adiabatic passage)~\cite{deiglmayrCoherentExcitationRydberg2006} between basis states. Figure~\ref{fig:SOAT_TIRAP} shows the result of the application of such a pulse sequence to an atom initially prepared in one of the basis states. The fitting routine identifies the initial state and final state with an error of less than \SI{5}{\percent}, in the transition regime the transient population in the intermediate state is detected. \\

  \textit{Tunable couplings and on-site energies.} To faithfully realize the system described by Hamiltonian~\eqref{eq:SOAT_rice-mele} within a manifold of Rydberg states, the Rabi frequencies $\Omega_{ij}$ and detunings $\Delta_{ij}$ of the rf fields coupling the basis states $i$ and $j$ need to be controlled accurately. Also, couplings to other states outside of the chosen basis need to be  suppressed. To this end, a magnetic field of \SI{17}{\gauss} is applied during Rydberg excitation and application of rf fields, which splits the Zeeman-sublevels of the basis states. The pyramidal horn is oriented with the electric field plane perpendicular to the magnetic field direction such that the rf field drives predominantly $\sigma^+$ and $\sigma^-$ transitions.
  With the choice of the magnetic sub levels $m_j=1/2$ and $m_j=3/2$ for the ${\vphantom 1}^2\mathrm{S}_{1/2}$ and ${\vphantom 1}^2\mathrm{P}_{3/2}$ basis states, respectively, the detuning of the nearest dipole allowed transition coupling to the outside of the system, ${\vphantom 1}^2\mathrm{S}_{1/2}, m_j=1/2 \rightarrow {\vphantom 1}^2\mathrm{P}_{3/2}, m_j=-1/2$, is about \SI{60}{\mega\hertz}. The Rabi frequencies $\Omega_{ij}=2 J_{ij}$, which are twice the couplings $J_{ij}$ between sites $i$ and $j$ in the synthetic dimension, are calibrated \textit{in-situ} using Autler-Townes spectroscopy~\cite{autlerStarkEffectRapidly1955} (see appendix~\ref{sec:autlertownes}). In order to implement simultaneous coupling of many states, the AWG is programmed with a waveform that is the sum of the required oscillating fields with time-dependent amplitudes and frequencies~(see appendix~\ref{sec:waveforms}).

\begin{figure}[tb!]
	\centering
    \includegraphics[width=1.\linewidth]{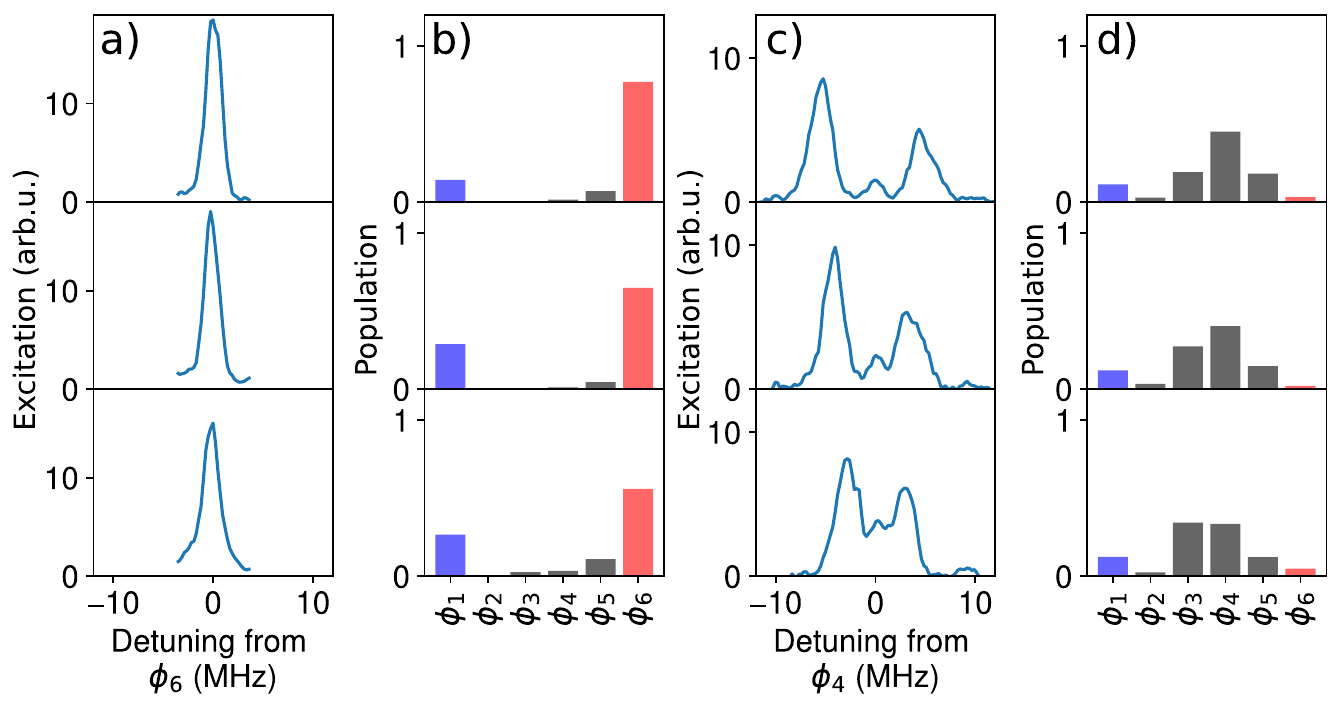}
	\caption{\label{fig:SOAT_exp_SSH} Probing the SSH Hamiltonian in a system of six coupled Rydberg states with $J_1=J_W=2\pi\times\SI{1}{\mega\hertz}$ with an UV pulse of $\SI{3}{\micro\second}$ length. (a) Total number of Rydberg excitations, when the UV laser frequency is close to the edge state $\phi_6$, for different couplings strengths $J_2=J_S=2\pi\times \{\SI{2}{\mega\hertz}, \SI{3}{\mega\hertz}, \SI{4}{\mega\hertz}\}$ (from bottom to top). (b) Measured populations in the basis states at the maximum of the Rydberg signal in a). (c) As panel a), but with the UV laser frequency close to the bulk state $\phi_4$. (d) Averaged populations in the basis states measured at the two maxima of the Rydberg signal in c).}
\end{figure}

  For $\Delta=0$ and time-independent couplings $J_1 < J_2$ (in the following referred to as ``weak'', $J_W$, and ``strong'', $J_S$, couplings, respectively) the RM becomes equivalent to the SSH model. For the open boundary conditions of the implemented systems, the spectrum of the SSH Hamiltonian in the thermodynamic limit consists of two bands comprising excitations of the bulk, and two topologically-protected edge states at zero energy in the middle of the two bands~\cite{suSolitonsPolyacetylene1979,kanungoRealizingTopologicalEdge2022}. We probe this structure by turning on the rf fields first, followed by a UV laser pulse and state-selective detection. If the laser is close to resonance with a Rydberg state at the edge of the system, only a single peak at zero detuning from the isolated atomic resonance is observed [Fig.~\ref{fig:SOAT_exp_SSH}\, a)]. The state-selective detection reveals indeed contributions from both edge states to this zero-energy resonance [Fig.~\ref{fig:SOAT_exp_SSH}\, b)]. With increasing ratio of the couplings strengths $J_S/J_W$, the contribution from the second edge state, located at the other end of the chain, diminishes. We attribute this decoupling to residual ac-Stark shifts which could be independently measured and compensated for in subsequent studies~\cite{luProbingTopologicalPhase2024}. If the laser is close to resonance with a Rydberg state in the bulk of the system, a triplet of peaks is observed [Fig.~\ref{fig:SOAT_exp_SSH}\, c)]. The central peak at zero detuning, attributed to the edge states, quickly decreases in amplitude with increasing ratio of $J_S/J_W$, while the two outer peaks, corresponding to the two bulk bands, split by approximately the value of the strong coupling $J_S$. Probing the character of the excited state via state-selective ionization reveals indeed dominant contributions from the bulk states [Fig.~\ref{fig:SOAT_exp_SSH}\, d)].  We conclude that for couplings $J_1$, $J_2$ smaller than $2\pi\times\SI{4}{\mega\hertz}$, the experimentally realized system can faithfully reproduce the Rice-Mele model of Equation~\eqref{eq:SOAT_rice-mele}.

  \textit{Quantum transport in a synthetic dimension.} Having established the realisation of the RM Hamiltonian in our system, we are now ready to simulate the pumping of a quantum particle by periodic variation of the Hamiltonian's parameters. We modulate couplings and on-site energies in a way analogous to the Thouless' pump scheme~\cite{citroThoulessPumpingTopology2023}, namely, if the parameters would change infinitesimally slowly and in the limit of an infinite chain, one expects that after one period of the cyclic change of parameters depicted in Fig.~\ref{fig:SOAT_schematic_thouless}\,a), the initial Hamiltonian is restored, but the quantum particle has been pumped to the next unit cell. However, our system differs in two important ways from the Thouless pump: first, it is a short chain of three unit cells, and second we will not restrict ourselves to the adiabatic limit of infinitely long pumping periods, but instead investigate the pumping efficiency as a function of the pumping period. For optimal site resolution, we change the decomposition of the synthetic dimension into cells which are now formed by states $\{2,3\}$, $\{4,5\}$, and $\{6\}$ (compare Fig.~\ref{fig:SOAT_exp_tof_traces}). To initialize the system, an rf field is applied which sets $J_2 > J_1 = 0$ and $\Delta$=0. In this configuration, the system dimerizes and the coupling in each cell is identical to the situation discussed above in the context of the Autler-Townes effect. The new eigenstates of each cell are linear superpositions of the excitations at the two sites with eigenenergies $\pm J_2$. By tuning the Rydberg excitation laser  (UV pulse length \SI{2}{\micro\second}) to one of the components of the Autler-Townes doublet formed by $\phi_2$ and $\phi_3$, we prepare the initial state of the system as an eigenstate localized in cell 1. After this initialization, we perform a sequence of two pumping cycles (due to the limited size of the system) by playing a preprogrammed waveform on the AWG. After this sequence we state-selectively detect the populations. We characterize the success of pumping by measuring the \textit{transfer efficiency}, which we define as the ratio of the population in unit cell 3 over the total detected signal. In this implementation with five sites, cell 3 consists of state $\phi_6$ only, which minimizes detection errors because its time-of-flight signature is well separated from the ones of the other states (see Fig.~\ref{fig:SOAT_exp_tof_traces}).

\begin{figure}[tb]
	\centering
    \includegraphics[width=\linewidth]{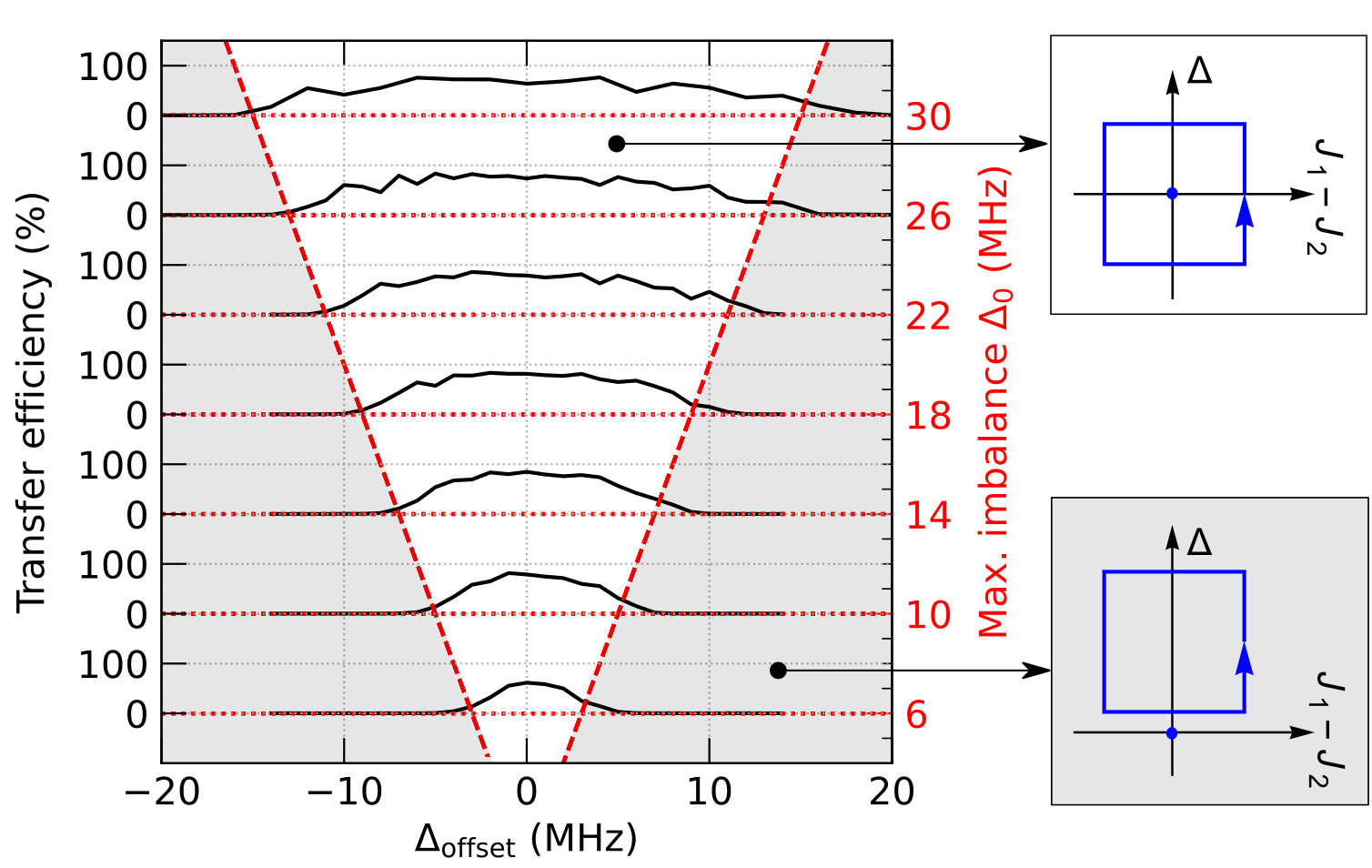}
	\caption{\label{fig:SOAT_exp_delta_assym} Dependence of the transfer efficiency (black curves and axis), defined as in the text, on the system's trajectory in parameter space for a fixed maximum coupling of  $J_{1,0}=J_{2,0}=2\pi\times\SI{2.5}{\mega\hertz}$ (two pump cycles with a pump period of \SI{1.25}{\micro\second}). $\Delta_\mathrm{offset}$ is an offset added to the time-dependent imbalance of the on-site energies, $\Delta(t)$. Curves for different amplitudes $\Delta_0$ of the imbalance's variation are vertically offset (dotted-red lines and axis). The dashed lines enclose the ``topological pumping'' region where the parameter trajectory encircles the band-gap-closing point (upper inset), the gray-shaded region marks the ``trivial'' region (lower inset).}
\end{figure}

    Thouless pumping is a topological effect: the number of charges pumped per cycle corresponds to a quantized Chern-number associated with the 2D torus defined by the 1D Brillouin zone and the parameter periodically cycled in the Hamiltonian~\cite{ozawaTopologicalQuantumMatter2019a}. In the RM model, the origin of the parameter space is a singular point of the Berry curvature where the band gap closes, and the Chern number is proportional to the number of windings the trajectory takes around this singular point. To examine experimentally whether such topological behavior is present in our pump, we vary the trajectory depicted in Figure~\ref{fig:SOAT_schematic_thouless}\,a) and translate it in parameter space by a variable offset of the imbalance of the on-site energies~(appendix~\ref{sec:optimalperiod}). The experimentally measured dependence of the transfer efficiency on these variations is shown in Figure~\ref{fig:SOAT_exp_delta_assym}: one observes a nearly constant transfer efficiency on the order of \SI{90}{\percent} as long as the parameter's trajectory encloses the origin, while the efficiency quickly drops to zero if the origin is not encircled. Therefore, we clearly see the existence of two pumping regimes: a ``topological'' pumping regime where the cycle encloses the origin of parameter space and transfer efficiencies reach values of up \SI{90}{\percent}, and a ``trivial'' pumping regime when the parameters do not enclose the origin and transport becomes negligible.

\begin{figure}[tb]
	\centering
    \includegraphics[width=1.0\linewidth]{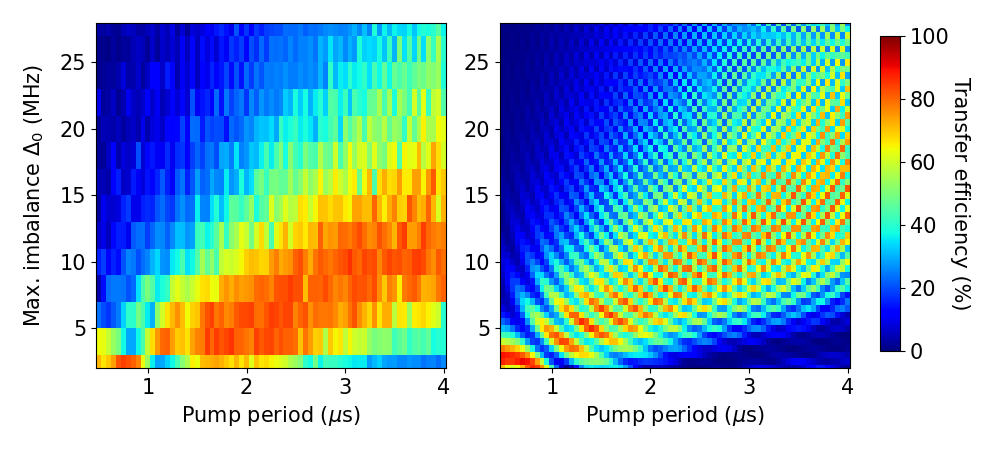}
	\caption{\label{fig:SOAT_exp_theory_efficiency} Left panel: transfer efficiency (color coded) \textit{vs} pump-cycle period and maximum detuning $\Delta_0$ of the parameter's variation. Transfer efficiency is defined as the ratio of the population in the state $\phi_6$ over the total detected signal at the end of the sequence. The maximum couplings $J_{1,0}=J_{2,0}$ are $2\pi\times\SI{1.5}{\mega\hertz}$. Right-hand panel: Simulated dynamics for the chosen experimental parameters (see text for details). The simulated transfer efficiencies exhibit highly oscillatory variations which are only weakly observable in the experimental data, which can be attributed to inhomogeneities in the parameters over the atomic sample or dephasing during the evolution of the wavefunction.}
\end{figure}

    \textit{Adiabaticity of the pump.} We now discuss the impact of the rate of pumping and its adiabaticity on pumping efficiency. The left-hand panel of Figure~\ref{fig:SOAT_exp_theory_efficiency} depicts the transfer efficiency as function of the pump-cycle period and the maximum detuning $\Delta_0$ for a fixed maximum coupling strength $J_{1,0}=J_{2,0}= 2\pi\times\SI{1.5}{\mega\hertz}$. For a given $\Delta_0$, the transfer efficiency first increases with increasing pump period, which can be attributed to increasing adiabaticity of the dynamics (in the sense of Landau-Zener's theory of level crossings~\cite{zenerNonadiabaticCrossingEnergy1997}) as the parameters' rate of change is reduced. Surprisingly, we observe that the efficiency reaches a maximum at certain pumping period, beyond which it decreases monotonically (\textit{e.g.} in Figure~\ref{fig:SOAT_exp_theory_efficiency} for a $\Delta_0$ of $\SI{4}{\mega\hertz}$ and pump periods exceeding $\SI{3}{\micro\second}$) which at first appears puzzling since the ideal Thouless pump is expected to operate better as the pumping becomes more adiabatic. To gain further insights, the Schrödinger equation of Hamiltonian~\eqref{eq:SOAT_rice-mele} is solved numerically~(appendix~\ref{sec:optimalperiod}). The main characteristics of the experimental observations are well captured by a simulation of the single-particle dynamics under the programmed change of parameters shown on the right-hand side of Figure~\ref{fig:SOAT_exp_theory_efficiency}. Our numerical studies indicate that the observed decrease of the transfer efficiency is not associated with a decrease of the ability of the pump to transport the particle according to the ideal quantized expectation (\textit{i.e.}, one unit cell per period), but rather to a combination of the effects of the spread of the particle wave-function and our limitations to fully measure its position. Namely, our simulations show that even for long pump periods the pump is still able to move the mean position of the particle according to the expected ideal quantized motion. The dispersing wave-function is, however, reflected at the ends of the chain because of the open boundary conditions, which prevents us from extracting the translation of the mean position from the occupation of sites~(see appendix~\ref{sec:optimalperiod}).

    \textit{Conclusions.} In this letter, we have demonstrated the implementation of the RM Hamiltonian in a synthetic dimension of the Rydberg states of cesium atoms and the realization of topological Thouless pumping, as summarized in Fig.~\ref{fig:SOAT_exp_delta_assym}, where we see that when our pump follows a trajectory in Hamiltonian parameter space that encloses the Berry curvature singularity located at the origin, there is an efficiency of about \SI{90}{\percent} for transporting the particle by one unit cell per cycle along the synthetic dimension, and when the pumping loop does not enclose this Berry curvature singularity we observe a rapid drop of the transfer efficiency towards negligible values.

    There are several interesting directions for future research that we anticipate from our study, such as the implementation of driven systems with periodic boundary conditions~\cite{shenSimulatingQuantumMechanics2022,chenQuantumWalksCorrelated2024}, non-Abelian properties~\cite{youObservationNonAbelianThouless2022}, or strong interactions~\cite{jurgensenQuantizedNonlinearThouless2021,fengQuantumMembranePhases2022,walterQuantizationItsBreakdown2023}. In addition, our system could serve as a new platform to implement a variety of dynamical regimes that are harder to investigate in more traditional many-body settings, such as the establishment of late-time non-equilibrium steady states of dynamically driven and quantum systems \cite{okaFloquetEngineeringQuantum2019,geierFloquetHamiltonianEngineering2021}.

\begin{acknowledgments}
 We thank Prof. Jan Meijer, Leipzig University, and his group for lending the AWG to us, and Lorenz Linnemann for perfoming additional numerical simulations. This work was supported by the Deutsche Forschungsgemeinschaft through SPP 1929 (GiRyd) under Project No. 428456632.
\end{acknowledgments}

\appendix

\section{Calibration of coupling strengths}
\label{sec:autlertownes}

\begin{figure}[b!]
    \includegraphics[width=0.85\linewidth]{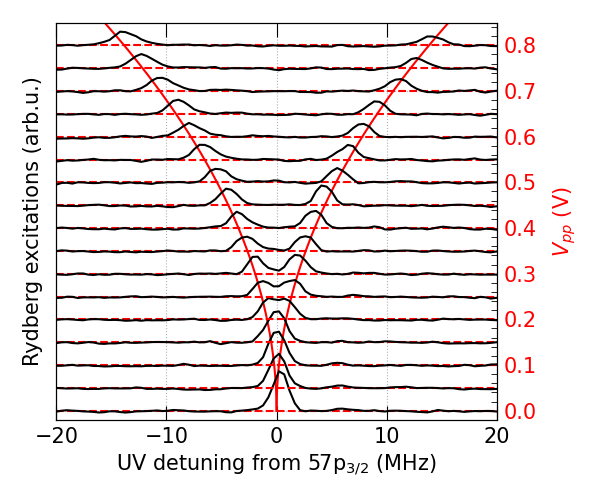}
       	\caption{\label{fig:SOAT_exp_autler_townes} (black traces) Rydberg excitations as a function of the detuning of the UV laser from the transition $\phi_6\leftarrow 6^2\mathrm{S}_{1/2}(F=4,m_F=4)$ in the presence of an rf field coupling the states $\phi_6$ and $\phi_5$ at \SI{21451.68}{\mega\hertz}. The traces are offset by the peak-to-peak voltage $V_{pp}$ of the rf tone set in the AWG (red-dashed lines and right-hand vertical axis). The very weak resonance at \SI{+6}{\mega\hertz} results from a residual population in $6^2\mathrm{S}_{1/2}(F=4,m_F=3)$ after optical pumping. (Red curve) fit of a parabola to the peak positions determined in separate fits (not shown).}
\end{figure}

    The Rabi frequencies $\Omega_{ij}=2 J_{ij}$, which are twice the couplings $J_{ij}$ between sites $i$ and $j$ in the synthetic dimension, are calibrated \textit{in-situ} using Autler-Townes spectroscopy~\cite{autlerStarkEffectRapidly1955}. As an example, the calibration measurement for the transition $\phi_5\rightarrow \phi_6$ is shown in Figure~\ref{fig:SOAT_exp_autler_townes}: the Autler-Townes splitting, which equals the Rabi-Frequency of the rf-atom coupling, is measured for several amplitudes $V_{pp}$ programmed into the AWG. The passive frequency doubler causes the electric-field amplitude at the location of the atoms, and thus the Rabi frequency, to scale quadratically with the amplitude $V_{pp}$ of the field created by the AWG. We observe the scaling $\Omega_{ij} = \alpha_{ij} V_{pp}^2$ for Autler-Townes splittings of up to \SI{30}{\mega\hertz} and extract the coefficients $\alpha_{ij}$ by non-linear regression. From the broadening of the Autler-Townes peaks we estimate an inhomogeneity of the Rabi frequencies over the atomic sample of about \SI{10}{\percent}, consistent with the observed loss of contrast in Rabi-cycling measurements after a few cycles (not shown). We attribute this inhomogeneity to the distribution of rf field amplitudes over the atomic sample, which has a typical diameter (full-width-at-half-maximum) of \SI{1}{\milli\meter}.

\section{Calculation of rf waveforms}
\label{sec:waveforms}

In order to implement simultaneous coupling of many states, the AWG is programmed with a waveform that is the sum of the required oscillating fields with time-dependent amplitudes and frequencies. The individual amplitudes are scaled such that the maximum total amplitude equals the full scale of the AWG, and the output amplitude of the AWG (option AC) is configured to realize the chosen atomic-coupling strengths. The employed AWG has a resolution of 10 bit, \textit{i.e.} roughly 8 bit per rf frequency for a system of six coupled sites. Because of the use of a frequency doubler, the programmed amplitudes scale with the square root of the desired Rabi frequencies. This scaling turns out to be advantageous, because it reduces the differences in the amplitudes of the field components and distributes quantization noise more evenly. As a concrete example, the programmed relative amplitudes $v_{ij}$ to achieve identical couplings $\Omega_{ij}$ between the basis states $\phi_i$ and $\phi_j$ are $v_{12}=\SI{27}{\percent}$, $v_{23}=\SI{18}{\percent}$, $v_{34}=\SI{22}{\percent}$, $v_{45}=\SI{18}{\percent}$, and $v_{56}=\SI{15}{\percent}$ of the full scale of the AWG.

\begin{figure}[tb!]
	\centering
    \includegraphics[width=0.9\linewidth]{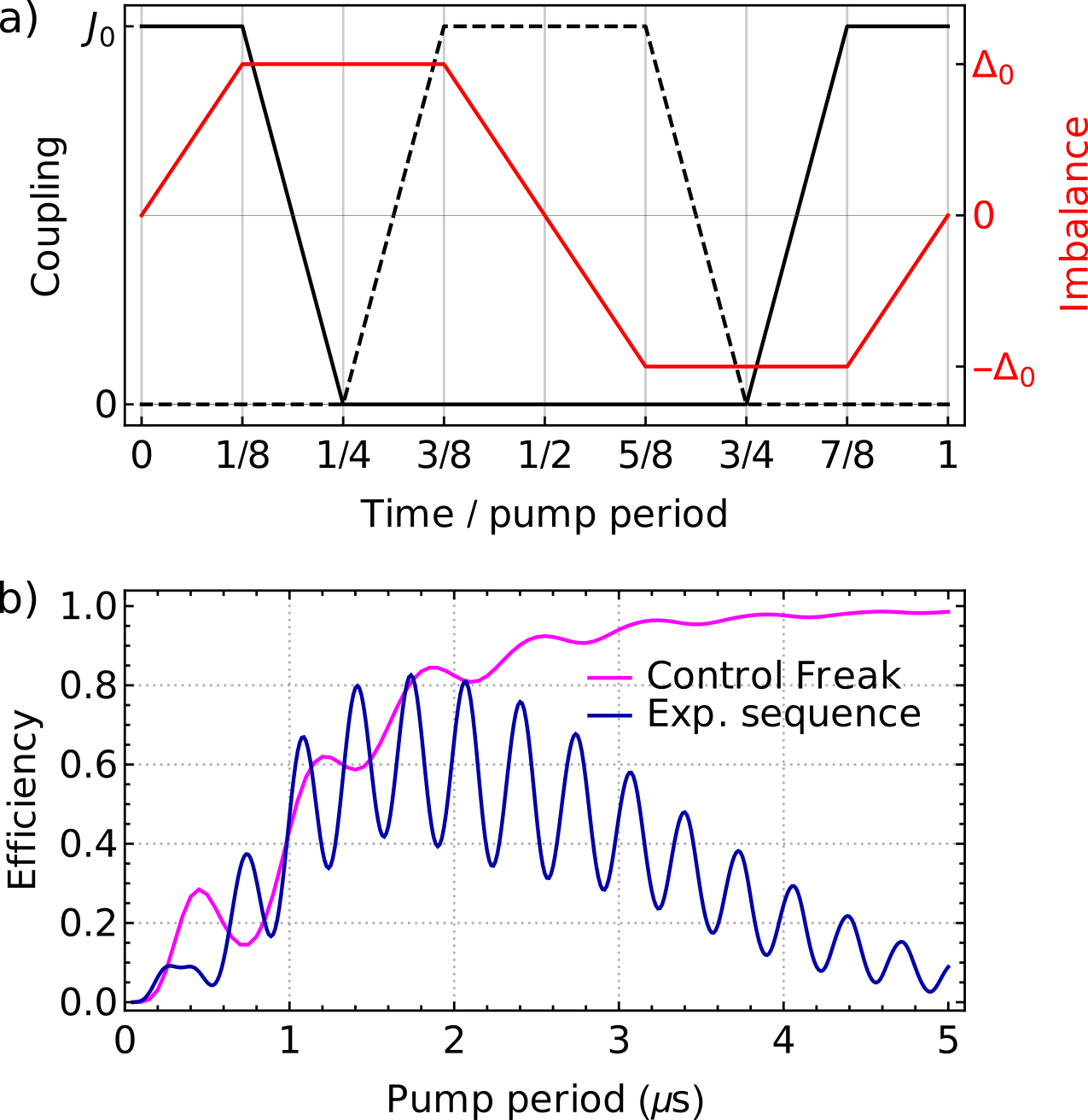}
	\caption{\label{fig:app_controlfreak} a): Time dependence of the intra-dimer couplings $J_1$ (black), inter-dimer couplings $J_2$ (dashed), and imbalance of on-site energies $\Delta$ (red) in the ``Control Freak'' parameterization~\cite{asbothAdiabaticChargePumping2016}. b) Transfer efficiency after two pump cycles as defined in the main text for a system size of $N=5$, maximum couplings $J_{1,0}=J_{2,0}=\Omega_0=2\pi\times\SI{1.5}{\mega\hertz}$, and maximum detuning $\Delta_0=2\pi\times\SI{7}{\mega\hertz}$ for different periods of the pump cycle: (magenta) for the ``Control freak'' parameterization depicted in panel a), and (blue) for the parameterization used in the experiments [see also Fig.~\ref{fig:app_bands}\,a)].}
\end{figure}

\section{Optimal pump-cylce period}
\label{sec:optimalperiod}

In this section we investigate the mechanism behind the optimal pump period of the pumping protocol. Having Landau and Zener's theory of the dynamics at level crossings in mind, one might expect that the slower the parameters of the model are varied, the more efficient the pumping will be due to enhanced adiabaticity~\cite{zenerNonadiabaticCrossingEnergy1997}. However, we observe experimentally, see Fig.~6 of the main text, that the efficiency grows to a maximum at some pump duration (which we call the optimal pump-cycle period), and then decreases for longer durations of the pump pulse. We will demonstrate that this is neither an effect of the finite system size, nor of decoherence. The key mechanism behind this optimal pump period is the existence of a finite band-width of the states in our model that leads to a quantum mechanical delocalization of the state in the synthetic dimension. To illustrate the origin of this mechanism, we compare numerical simulations of the system's dynamic (using the Mathematica package AtomicDensityMatrix, Rochester Scientific) for two different ways to vary the parameters of the model.

In both cases, the Thouless pumping protocol, depicted in Fig.~1\,a) of the main text, is realized. In scenario 1, shown in Fig.~\ref{fig:app_controlfreak}\,a), the so-called ``control freak'' parameterization~\cite{asbothAdiabaticChargePumping2016} is applied. At any point in the cycle, either $J_1$ or $J_2$ is zero. Thus the chain is always dimerized, corresponding to a ``flat band'' situation. As shown in Figure \ref{fig:app_controlfreak}\,b), the pumping efficiency in this scenario raises almost monotonically, which can be attributed to the increasing adiabaticity of the transfer process when the parameters of the Hamiltonian are varied more slowly.

\begin{figure*}[tb!]
    \includegraphics[width=0.95\textwidth]{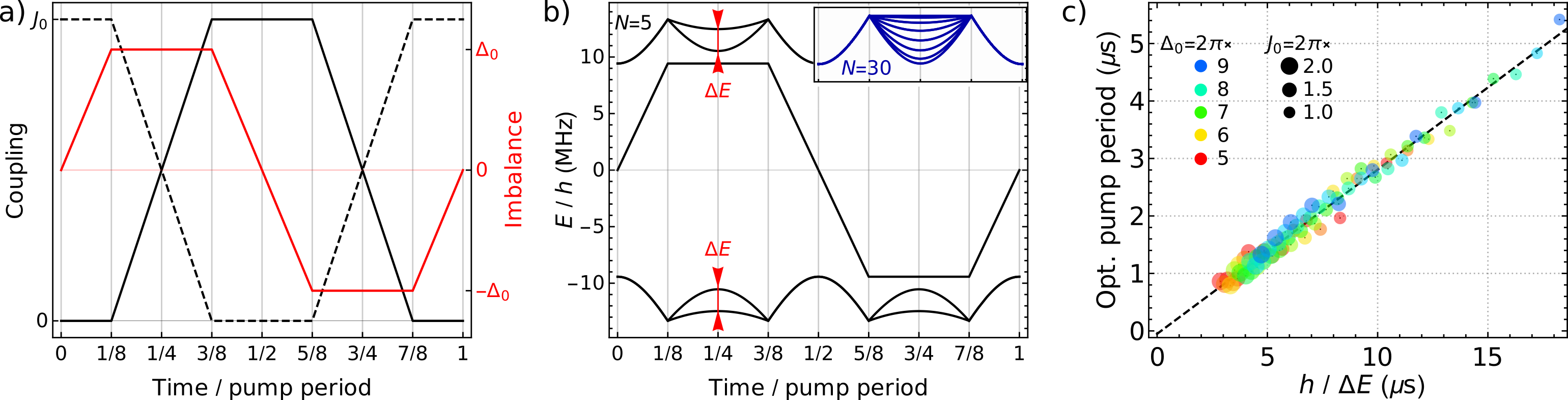}
	\caption{\label{fig:app_bands} a): Time dependence of the intra-dimer couplings $J_1$ (dashed), inter-dimer couplings $J_2$ (black), and imbalance of on-site energies $\Delta$ (red) for one cycle in parameter space as set in the experiments. b) Instantaneous spectrum of the RM Hamiltonian for a system size of $N=5$, maximum couplings $J_{1,0}=J_{2,0}=J_0=2\pi\times\SI{1.5}{\mega\hertz}$, and maximum detuning $\Delta_0=2\pi\times\SI{3}{\mega\hertz}$. The inset depicts the instantaneous spectrum for a system size of $N=30$, and the red arrows mark the width of the band $\Delta E$ used to present the data in panel c). c) Pump-cycle period for which optimal transfer is reached in a system of size $N=5$, versus the width of the band for numerical simulations with $J_{1,0}=J_{2,0}=J_0$ varying between $2\pi\times\SI{1}{\mega\hertz}$ and $2\pi\times\SI{2}{\mega\hertz}$, and $\Delta_0$ varying between $2\pi\times\SI{5}{\mega\hertz}$ and $2\pi\times\SI{9}{\mega\hertz}$. Color and size of the points indicate the specific values of $\Delta_0$ and $\Omega_0$ (see legend) for which the optimal pump period was determined.}
\end{figure*}

\begin{figure}[tb!]
	\centering
    \includegraphics[width=0.9\linewidth]{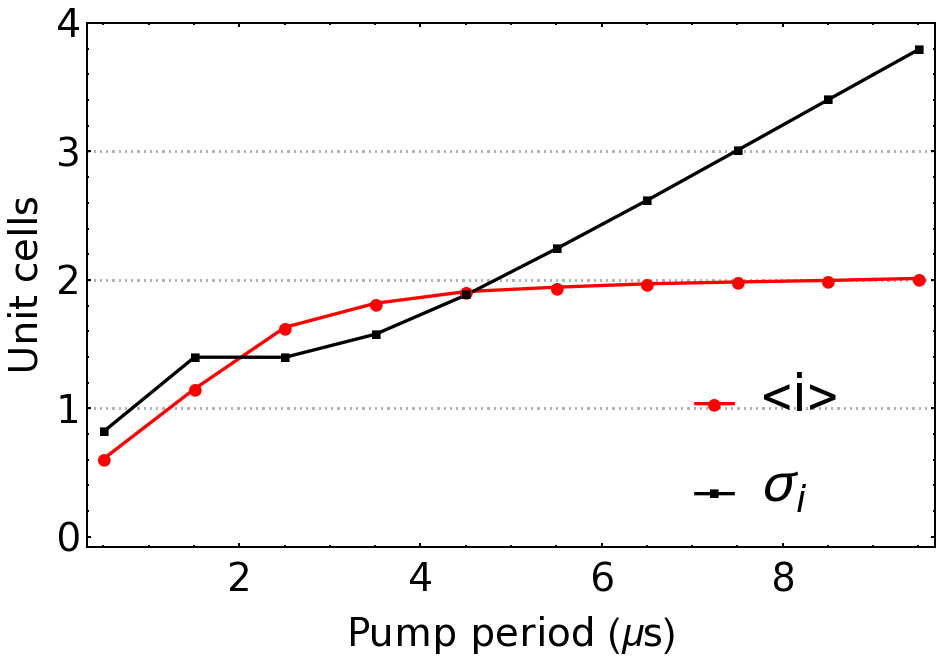}
	\caption{\label{fig:app_mean_position} Numerical simulation of the shift of the mean particle position $\langle i \rangle$ (where $i$ labels unit cells) and the spread of the wave-function  $\sigma_i$ (one standard deviation) after two cycles of the pumping protocol used in the experiments [see Fig.~\ref{fig:app_bands}\,b)] with maximum couplings $J_{1,0}=J_{2,0}=2\pi\times\SI{1.5}{\mega\hertz}$ and maximum imbalance of on-site potentials $\Delta_0=2\pi\times\SI{8}{\mega\hertz}$ for different lengths of the pump period. Initially the particle is located in the center of long chain consisting of 15 unit cells, such that edge effects are negligible.}
\end{figure}

In scenario 2, depicted in Fig.~\ref{fig:app_bands}\,a) and realized in the experiments described in the main text, the pumping efficiency exhibits a maximum at intermediate pump period before decreasing again for a longer pump period. To develop an intuitive explanation for the existence of this optimal pump period in a decoherence-free simulation, we investigate more closely the time in the pumping protocol when all sites are coupled with equal strength, \textit{i.e.} $t=1/4$ or $3/4$ in Fig.~\ref{fig:app_bands}\,a). In this situation, highlighted by the red arrows in  Fig.~\ref{fig:app_bands}\,b), the degeneracies of the dimerized configuration are lifted and the instantaneous eigenstates form a band with a non-zero band width, in contrast to the ``control-freak'' protocol where the instantaneous band-width is always zero. For a finite system, the energies in these bands are still clearly discretized, but with increasing system size they approach a continuum, as demonstrated for $N=30$ in the inset of Fig.~\ref{fig:app_bands}\,b). The well known dispersion relation for propagation of localized wavepackets in an energy band states that, for long times, the width of the wavepacket increases linearly in time where the coefficient is given by the effective mass obtained from the curvature of the bottom of the band in momentum space~\cite{kittelIntroductionSolidState1996}. This is apparent in the particle's mean position and the uncertainty in its position, which are shown in Fig.~\ref{fig:app_mean_position} under pumping with the experimental parameterization [Fig.~\ref{fig:app_bands}\,a)], but in a much larger chain where edge effects are negligible.

The curvature of the band structure is proportional to the band width. In Fig.~\ref{fig:app_bands}\,c), we plot the simulated optimal pump-cycle period versus the (inverse) maximum width of the bands for different maximal coupling strengths $J_{1,0}=J_{2,0}$ and on-site energy imbalances $\Delta_0$. Strikingly, the  points collapse onto a single curve, revealing a strong correlation between the optimal pump period and the inverse width of the band, \textit{i.e.}, the inverse of the velocity with which an initially localized wavepacket disperses in the given band structure. For large system sizes ($N \gtrsim 15$), the optimal pump-cycle period $t_\mathrm{opt}$ approaches $t_\mathrm{opt} \approx \sfrac{h}{\Delta E}$ where $\Delta E$ is the maximum width of a single band [see Fig.~\ref{fig:app_bands}\,b)]. The following intuitive picture is thus consistent with our experimental observations and the numerical simulations: the pumping efficiency first increases with increasing pump period, because the pump dynamics becomes more adiabatic. For longer times, however, the excitation, initially localized on a single dimer, spreads over several dimers, resulting in a reduction of the observed pumping efficiency.

\begin{figure}[bh]
	\centering
    \includegraphics[width=0.95\linewidth]{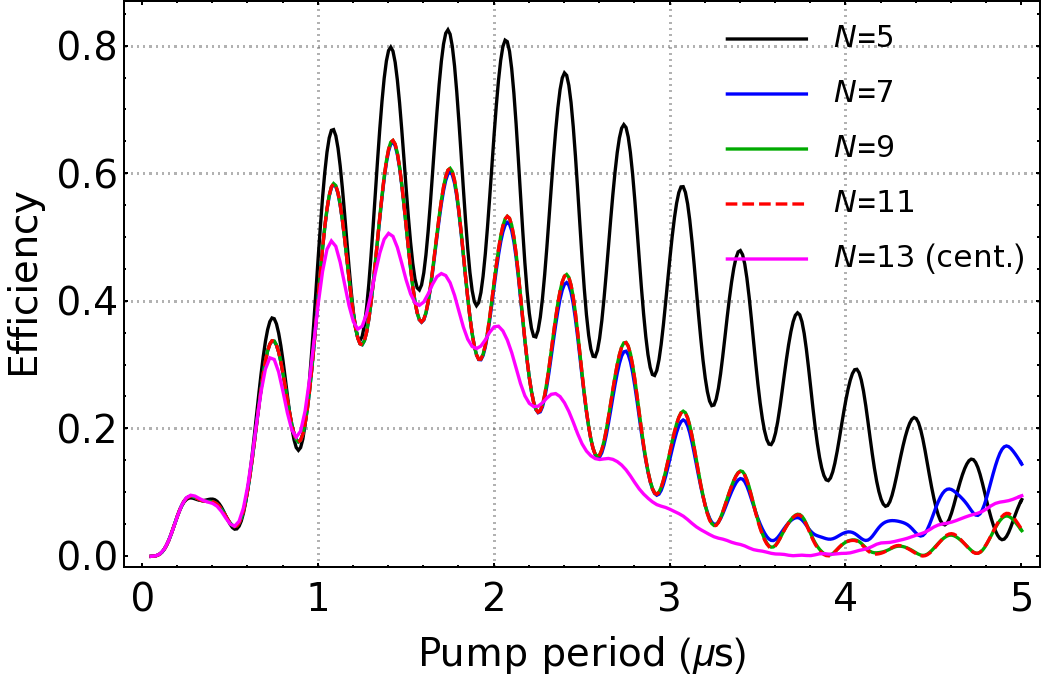}
	\caption{\label{fig:app_sizedep} Transfer efficiency as defined in the main text for different system sizes, maximum couplings $J_{1,0}=J_{2,0}=\Omega_0=2\pi\times\SI{1.5}{\mega\hertz}$, and maximum detuning $\Delta_0=2\pi\times\SI{7}{\mega\hertz}$ for different lengths of the pump cycle. The particle is always initialized in cell 1, except for the case $N=13$, where it is initialized in the central cell.}
\end{figure}

We finally investigate if this interpretation also holds for larger systems. In Fig.~\ref{fig:app_sizedep}, we plot the transfer efficiency as a function of pump-cycle period for different system sizes. Even though finite-size effects clearly modify the system's response for $N=5$, the most prominent feature, \textit{i.e.} the existence of an optimal pump period, persists for larger $N$. We note that for larger pump periods and system sizes, the efficiency exhibits a ``rephasing-like'' behaviour, which we attribute to the reflection of the wave packet from the open boundary. We also note that the frequency of the fast oscillations observed in all curves depends only on the maximum imbalance of on-site energies $\Delta_0$ and can thus be attributed to fast oscillations of the populations within (instantaneous) dimers.

\end{document}